\documentclass[%
 aip,
 amsmath,amssymb,
 reprint,%
]{revtex4-1}
\usepackage{graphicx}
\usepackage{dcolumn}
\usepackage{bm}
\usepackage[utf8]{inputenc}
\usepackage[T1]{fontenc}
\usepackage{mathptmx}
\usepackage{etoolbox}
\makeatletter
\def\@email#1#2{%
 \endgroup
 \patchcmd{\titleblock@produce}
  {\frontmatter@RRAPformat}
  {\frontmatter@RRAPformat{\produce@RRAP{*#1\href{mailto:#2}{#2}}}\frontmatter@RRAPformat}
  {}{}
}%
\makeatother
\begin{document}
\preprint{AIP/123-QED}
\title[Broadband Resonance-Enhanced Frequency Generation by Four-Wave Mixing in a Silicon Floquet Topological Photonic Insulator
]{Broadband Resonance-Enhanced Frequency Generation by\\Four-Wave Mixing in a Silicon Floquet Topological Photonic Insulator}
\author{T. J. Zimmerling}
\email{tzimmerl@ualberta.ca}
\author{S. Afzal}%
\author{V. Van}
\homepage{http://nprl.ece.ualberta.ca/}
\affiliation{Department of Electrical and Computer Engineering, Faculty of Engineering, University of Alberta, Donadeo Innovation Centre for Engineering, 9211 116 Street NW, Edmonton, Alberta, Canada T6G 1H9}
\date{\today}
\begin{abstract}
Floquet topological photonic insulators, whose light transport properties are dictated by the periodic drive sequence of the lattice, provide more flexibility for controlling and trapping light than undriven topological insulators, which can enable novel nonlinear optics applications in topological photonics.  Here, we employ a novel resonance effect called Floquet Defect Mode Resonance in a 2D silicon Floquet microring lattice to demonstrate resonance-enhanced frequency generation by four-wave mixing of Floquet bulk modes in the presence of Kerr nonlinearity. The compact, cavity-less resonance mode, induced through a periodic perturbation of the lattice drive sequence, has the largest reported Q-factor for a topological resonator of $\sim10^5$ with low group velocity dispersion, which enables efficient broadband frequency generation over several Floquet-Brillouin zones of the Floquet topological insulator. We achieved wavelength conversion over 10.1 nm spectral range with an average enhancement of 12.5~dB in the conversion efficiency due to the Floquet Defect Mode Resonance.  Our work could lead to robust light sources generated directly on a topologically-protected photonic platform.
\end{abstract}
\maketitle
\section{Introduction\label{sec:Introduction}}
Floquet topological photonic insulators (TPIs) based on periodically-driven quantum systems provide an exotic platform for realizing photonic devices with unique properties. Exploring beyond the topological protection of the edge modes, it is possible to further control the flow of light through the lattice by perturbing the drive sequence of the system, which can lead to localized defect modes whose spatial intensity distributions are dictated by the hopping sequence of the lattice. In particular, we have recently shown\cite{Afzal2021} that by introducing a phase detune to part of the drive sequence of a 2D Floquet TPI microring lattice, we could excite a Floquet mode spatially localized in a loop pattern with quasi-energy located in a topologically-nontrivial bulk bandgap, leading to a resonance effect called Floquet Defect Mode Resonance (FDMR). Here, we exploit the periodicity of the quasienergy of the Floquet defect mode and its low group velocity dispersion in a 2D microring lattice to demonstrate resonance-enhanced four-wave mixing (FWM) over several Floquet-Brillouin (FB) zones. While nonlinear self-phase modulation effects have been investigated for soliton propagation in a 2D Floquet TPI based on periodically-coupled waveguide arrays,\cite{Mukherjee2020} our work represents the first investigation of parametric nonlinear interactions of Floquet bulk modes in a topological resonator with Kerr nonlinearity, leading to enhanced broadband frequency generation on a Floquet topological platform.
\par
Parametric processes based on third-order nonlinearity in a topological photonic platform have attracted considerable interest due to their potential applications in parametric amplification, wavelength conversion, frequency generation and correlated photon pair generation. One-dimensional arrays of silicon nanodisks emulating the generalized Su–Schrieffer–Heeger model have been used to obtain enhanced third-harmonic generation at telecom frequencies through Mie resonances and the topological localization of edge states.\cite{Kruk2019} Plasmonic metasurfaces consisting of 2D hexagonal arrays of nanoholes in graphene sheets under static magnetic fields have been demonstrated to form Chern insulator edge states, which were used to achieve FWM in the 13 THz frequency range.\cite{You2020} Spontaneous FWM using edge modes in a Chern insulator based on 2D lattice of silicon ring resonators has also been demonstrated for correlated photon pair generation.\cite{Mittal2018} It should be noted that in these works, the nonlinear processes occur via the propagation of edge modes in static topological insulators, which require long lattice lengths for efficient frequency generation.
\par
Due to the large field buildups in an optical resonator at the resonance frequencies, strong nonlinear interactions of the resonance modes can take place, leading to highly efficient frequency generation by FWM in compact resonators. The use of resonance to enhance the conversion efficiency (CE) of FWM processes has been demonstrated in various conventional resonators and waveguide platforms,\cite{Ferrera2008, Ferrera2009,Turner2008,Azzini2013} with reported CE's of -26~dB in high-Q Hydex-in-silica resonators,\cite{Ferrera2008,Ferrera2009} and -25~dB in silicon-on-insulator (SOI) resonators.\cite{Turner2008} In order to achieve maximum CE enhancement due to resonance, the pump, signal and idler waves have to be aligned with the cavity resonances while simultaneously satisfying the phase matching condition, \mbox{$2\beta_p =  \beta_s + \beta_i$}, where $\beta_p, \beta_s$ and $\beta_i$ are the propagation constants of the pump, signal and idler, respectively, in the resonator.  However, dispersion of the resonant modes causes a phase mismatch, defined as \mbox{$\Delta\beta = 2\beta_p - \beta_s - \beta_i$}, which increases with the separation between the signal and idler waves, causing a precipitous degradation in the CE.\cite{Agrawal2019} For a fixed pump frequency $\omega_p$, the phase mismatch can be estimated as \mbox{$\Delta \beta \approx -\beta_2(\omega_p) \Delta \omega$}, where \mbox{$\beta_2 = d^2\beta/d\omega^2$} is the group velocity dispersion (GVD) and $\Delta \omega$ is the spectral separation between the signal and the pump. Thus, a requirement for achieving broadband wavelength conversion is that the resonance modes must have low GVD.
\par
In this paper, we exploit the high Q-factor and low dispersion of FDMRs in a silicon Floquet microring lattice to achieve efficient broadband wavelength conversion by FWM of Floquet bulk modes. A unique feature of the FDMR is that it is cavity-less, since its spatial localization pattern is defined not by any lattice boundaries or interfaces but instead by the hopping sequence of a Floquet bulk mode. The lack of physical interfaces enables the FDMR to have a very high Q-factor, with values as high as $\sim 10^5$ for our silicon structure, which is the largest reported Q-factor for a topological resonator. Combined with the low group velocity dispersion of the Floquet bulk mode, we achieved broadband wavelength conversion over 10.1 nm (from signal to idler wave), with an average CE enhancement of 12.5~dB due to the resonance effect.  Our work could lead to efficient broadband and tunable wavelength conversion, parametric amplification and correlated photon pair generation directly on a topologically-protected photonic platform.
\section{FDMR in a Floquet TPI Lattice}\label{sec:Theory}
\begin{figure*}[tbp]
    \centering
    \includegraphics[width=1\textwidth]{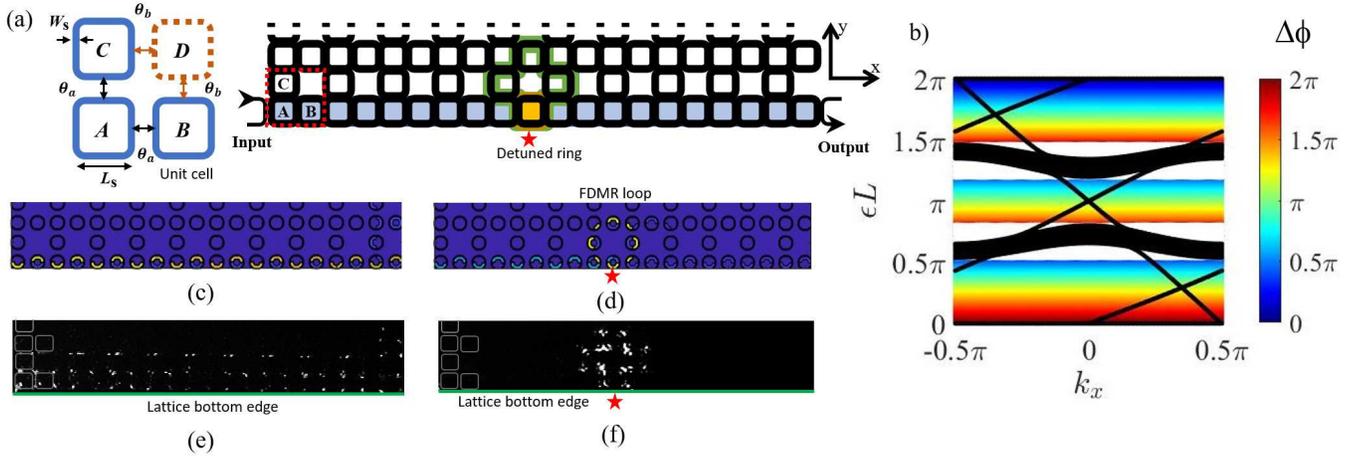}
    \caption{(a) Schematic of 2D Floquet microring lattice with each unit cell consisting of 3 identical microrings A, B, C with neighbor coupling angle $\theta_a$. The 4th microring D is omitted, setting $\theta_b\equiv0$. The FDMR loop, shown in green, is excited by applying a phase detune to 3/4 of the defect microring (indicated by red star). (b) Projected quasienergy band diagram of a Floquet lattice with $\theta_a=0.441\pi$ showing all 3 bandgaps hosting AFI edge modes.  The colored gradients depict FDMR quasienergy bands which move across the band gaps with increasing phase detune $\Delta\phi$ applied to the defect ring. (c) Simulated spatial distribution of an edge mode along the bottom lattice boundary at $\epsilon = 1.7\pi/L$, and (d) the induced FDMR at the same quasienergy. (e) NIR image of an edge mode and (f) induced FDMR at 1511.97~nm wavelength in the fabricated lattice.}
    \label{fig:TPISchematic}
\end{figure*}

Floquet Defect Mode Resonance can be regarded as the Floquet counterpart of point-defect resonance in static lattices, except that here the resonance mode arises from a periodic perturbation applied to the drive sequence of the Floquet lattice.\cite{Afzal2021} The Floquet TPI we consider here is a 2D square microring lattice which has been recently explored,\cite{Afzal2018,Afzal2020} with each unit cell consisting of four identical microrings characterized by 2 coupling angles ($\theta_a, \theta_b$) as shown in Fig.~\ref{fig:TPISchematic}~(a). The lattice exhibits Anomalous Floquet Insulator (AFI) behavior for \mbox{$\theta_a^2 + \theta_b^2 \gtrapprox \pi^2/8$}.  Here, we remove microring D by setting the coupling angle $\theta_b$ to 0, which helps reduce coupling loss and propagation loss in the lattice, allowing high-Q and strongly-localized FDMRs to be realized for enhanced nonlinear interactions. The resulting lattice, also shown in Fig.~\ref{fig:TPISchematic}~(a), is thus characterized by a single coupling angle $\theta_a$ between 3 resonators in each unit cell, and resembles a proposed AFI microring lattice.\cite{Liang2013}
\par
The Floquet nature of the lattice arises from the periodic circulation of light in each microring, with the driving Hamiltonian describing the sequence of couplings to neighbor resonators over one roundtrip (see supplementary material for details). We also note that unlike a similar microring lattice used to realize Chern insulator behaviour,\cite{Hafezi2013} our lattice does not use off-resonance link rings as couplers between site resonators, but instead employs direct evanescent coupling between neighbor resonators. This allows stronger coupling with lower dispersion to be achieved, so that our Floquet TPI exhibits nearly identical nontrivial band gaps over many FB zones, which is necessary for achieving broadband wavelength conversion by FWM.
\par
We designed our microring lattice to have coupling angle $\theta_a=0.441\pi$ so that it exhibits AFI behavior in all its three bandgaps over each FB zone.\cite{Afzal2018} Figure~\ref{fig:TPISchematic}~(b) shows the projected quasienergy band diagram, plotted as $\epsilon L$ where $\epsilon$ is the quasienergy and $L$ is the roundtrip length of the microrings, within one FB zone of a lattice with boundaries along the $x$-direction. Each FB zone also corresponds to one free spectral range (FSR) of the microrings. The black lines trace the quasienergies of the transmission bands and edge modes crossing three topologically nontrivial band gaps. The simulated intensity distribution of the edge mode at quasienergy $\epsilon$ = 1.7$\pi/L$ (located in band gap III) is shown in Fig.~\ref{fig:TPISchematic}~(c), which confirms light spatially localized along the bottom lattice boundary. 
\par
The FDMR is induced by applying a phase detune $\Delta\phi$ to the bottom three-quarters of a microring located on the bottom edge of the lattice (marked by a red star in Fig.~\ref{fig:TPISchematic}~(a)). The phase detune introduces a shift in the on-site potential of the defect resonator during three-quarters of each evolution period. This perturbation has the same periodicity of the driving sequence, since light experiences this phase detune each time it completes a roundtrip around the defect microring. The effect of the perturbation is to push a Floquet bulk mode from each transmission band into the band gap below, forming a flat-band state with a quasienergy shift proportional to the amount of phase detune. These flat bands are depicted by the horizontal color gradients in the band gaps in Fig.~\ref{fig:TPISchematic}~(b). The simulated spatial intensity pattern of the FDMR at quasienergy $\epsilon$ = 1.7$\pi/L$ is shown in Fig.~\ref{fig:TPISchematic}~(d), which shows light traveling in a loop pattern and constructively interfering with itself to form a strong resonance.\cite{Afzal2021} We note that this loop pattern is not due to light confinement by any physical lattice interfaces but follows the hopping sequence of a Floquet bulk mode in the microring lattice and extends far beyond the defect site resonator. Due to the periodic nature of the Floquet TPI, the FDMRs are also periodic in quasienergy, which allows resonance-enhanced FWM to be achieved over multiple FB zones. 
\section{Device Fabrication and Experimental Results}\label{sec:Setup}
\begin{figure}[tbp]
    \centering
    \includegraphics[width=0.48\textwidth]{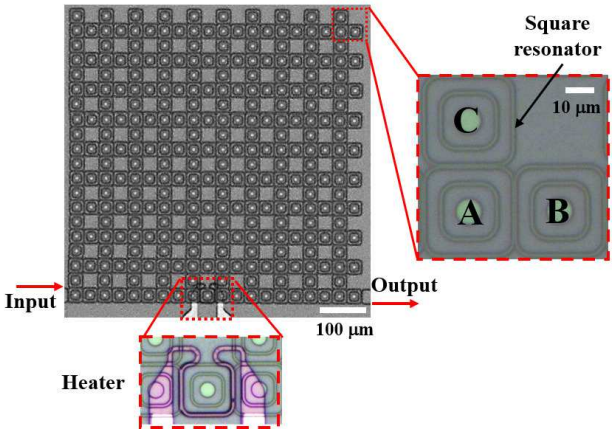}
    \caption{ Microscope image of the 10$\times$10 unit cell lattice with close-up images showing a unit cell and the metal heater on the defect microring.}
    \label{fig:Fabrication}
\end{figure}

We fabricated a Floquet microring lattice with 10$\times$10 unit cells on an SOI substrate, an image of which is shown in Fig.~\ref{fig:Fabrication}. The microring waveguides were designed for the fundamental TE mode with 450~nm width and 220~nm height, lying on a 2~$\mu$m-thick SiO$_2$ substrate and covered by a 3~$\mu$m-thick SiO$_2$ cladding. Each microring was designed to be a square with side lengths of 19.64~$\mu$m and rounded corners of 5 $\mu$m radius to reduce scattering loss. To achieve strong coupling necessary to realize the AFI phase, adjacent resonators were evanescently coupled along the side lengths through a coupling gap of 180 nm, providing a large coupling angle $\theta_a$ with low dispersion. A close-up image of a single unit cell is provided in Fig.~\ref{fig:Fabrication}. Simulations showed that the coupling angle remained relatively constant across multiple microring FSRs, with values of $0.431\pi$, $0.441\pi$ and $0.450\pi$ at the signal (1524.1~nm), pump (1529.1~nm), and idler (1534.2~nm) wavelengths, respectively, used to demonstrate FWM.
\par
To excite an edge mode along the bottom boundary of the lattice, an input waveguide was coupled to a microring at the bottom left corner and the transmitted light collected via an output waveguide coupled to a microring at the bottom right corner. A metallic heater with a resistance of $R\sim300~\Omega$ was used to apply a phase detune to three-quarters of a resonator on the bottom edge of the lattice via the thermo-optic effect, thereby generating an FDMR which is coupled to the AFI edge mode propagating along the bottom boundary.
\par
\begin{figure*}[htbp]
    \centering
    \includegraphics[width=0.7\textwidth]{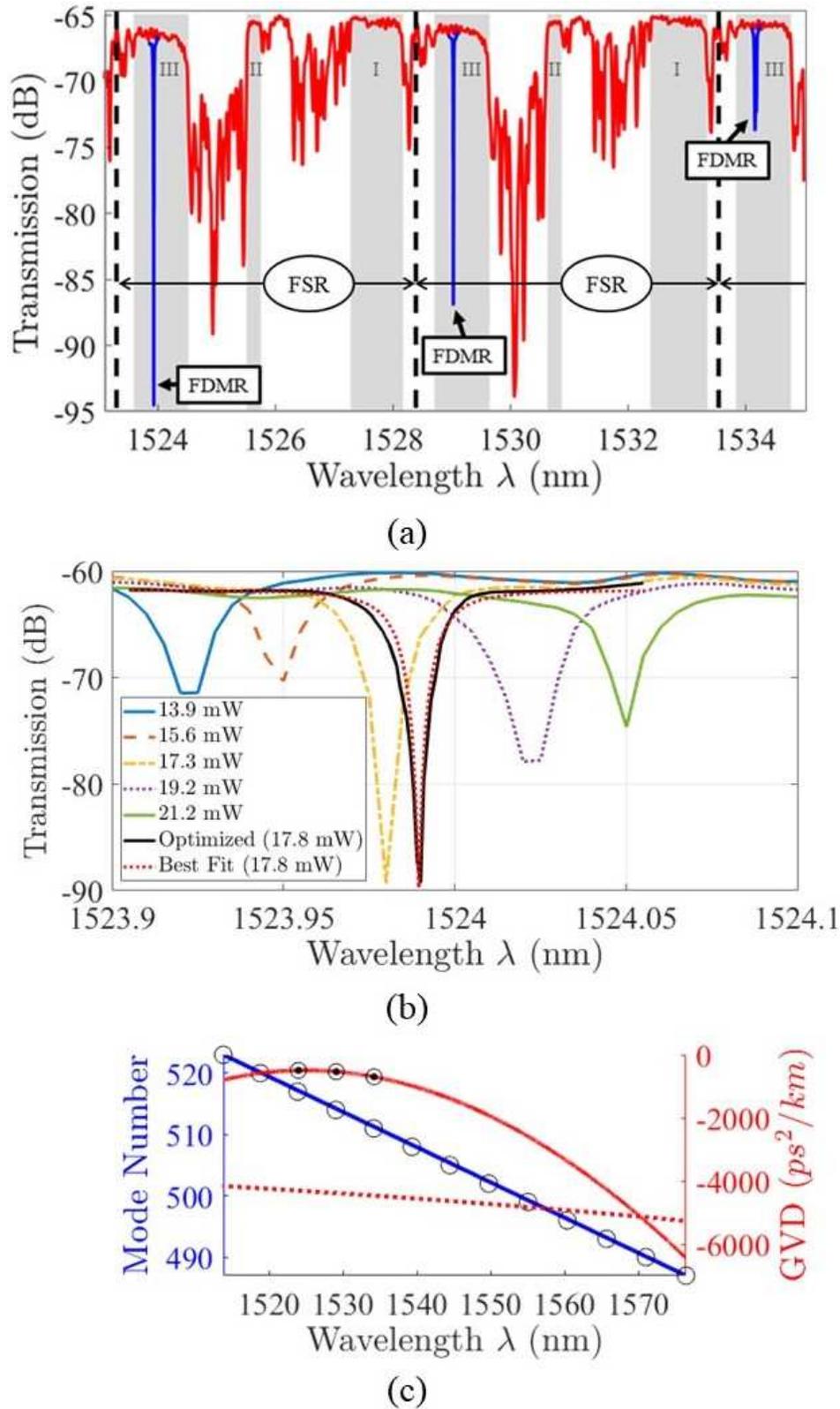}
    \caption{(a) Floquet TPI lattice spectrum in red with microring FSR marked by black dashes, FDMR in blue and lattice band gaps in gray. (b) FDMR resonance spectra with heating power from 13.9~mW to 21.2~mW, optimized FDMR achieving a Q-factor here of $7.26\times10^4$ with fitted curve. (c) FDMR resonance mode number versus wavelength with linear best fit line (blue). The group velocity dispersion (GVD) computed from experimental resonance wavelengths is shown by solid red line and the simulated dispersion is shown by the dotted line. The black data points at the signal, pump, and idler wavelengths provide GVD's ($\beta_2$) of -495\textpm2, -535\textpm2, and -697\textpm2 ps$^2$/km, respectively.}
    \label{fig:TPISpectra}
\end{figure*}

We first measured the transmission spectrum of the AFI edge mode by sweeping the wavelength of an input TE-polarized laser source with 3~$\mu$W of on-chip power over several FSRs of the microrings and measuring the transmitted power at the output waveguide (see supplementary material for details of the measurement setup). The result is presented by the red trace in Fig.~\ref{fig:TPISpectra}~(a), which shows three distinct wavelength regions in each FSR with relatively high and flat transmission due to edge mode propagation. We identify these regions as the three topological band gaps of the lattice, as predicted by the projected band diagram in Fig.~\ref{fig:TPISchematic}~(b). Figure~\ref{fig:TPISchematic}~(e) shows a near-infrared (NIR) image of the scattered light intensity of the lattice at 1511.97~nm wavelength, which confirms the presence of an edge mode in this band gap propagating along the bottom boundary. We next applied 17.8 mW of electrical power to the heater to induce a phase detune in the defect ring. The measured transmission spectrum at the output waveguide is presented by the blue trace in  Fig.~\ref{fig:TPISpectra}~(a), which is identical to the red trace except for the appearance of three sharp resonance dips. These dips correspond to an FDMR mode formed in band gap III of each microring FSR. Figure~\ref{fig:TPISchematic}~(f) shows the NIR image of the scattered light intensity taken at 1511.97~nm resonance wavelength with heating applied, which confirms the spatial localization of the resonance mode in a loop pattern. We note that the FDMR mode in band gap I is not visible due to weak coupling to the edge mode, while band gap II is too narrow to support a well-localized resonance mode.
\par
Figure ~\ref{fig:TPISpectra}~(b) shows the tuning of an FDMR across band gap III by varying the heater power from 13.9~mW to 21.2~mW. The linewidth of the resonance is also seen to vary across the band gap, which is due mainly to the dependence of the coupling between the edge mode and the FDMR loop on the roundtrip phase of the defect microring.\cite{Afzal2021} From the plot, we obtain the sharpest resonance dip near the mid gap at 17.8 mW heater power, yielding an FDMR with Q-factor of 72,600. As previously shown,\cite{Afzal2021} the FDMR also exhibits the strongest spatial localization as measured by its inverse participation ratio around the band gap center, which is important for achieving strong nonlinear effects. By fitting the resonance spectrum of the FDMR using an all-pass ring resonator model,\cite{VienBook} we extracted the effective coupling rate between the FDMR and the edge mode of $\mu=20.6$~GHz and an effective propagation loss of 4.54~dB/cm in the FDMR loop.
\par
From Fig.~\ref{fig:TPISpectra}~(a) we observe that the spectral distance separating the FDMR resonance dips is equal to the microring's FSR at 5.1~nm, although the roundtrip length of the FDMR loop is three times the microring circumference. This is due to the fact that the resonance modes in band gaps I and II are suppressed. Figure~\ref{fig:TPISpectra}~(c) plots the resonance mode number $m$ (blue circles) as a function of the measured resonant wavelength $\lambda_m$ from 1510~nm to 1580~nm. The resonance mode numbers are determined relative to the value $m = 514$ computed for the 1529.1~nm resonance from the formula \mbox{$m = 3L\cdot n_{eff}/\lambda_{m}$}, where $n_{eff}$ is the simulated effective index of the microring waveguide at 1529.1~nm and $3L = 329.9 ~\mu$m is the effective FDMR roundtrip length. We observe that the FDMR exists over a broad wavelength range with nearly constant mode spacing, as indicated by the best linear fit line.  To estimate the GVD of the FDMR mode, we first calculated its propagation constant \mbox{$\beta(\lambda_m)=2\pi n_m/\lambda_m$}, where the effective index $n_m$ is computed from the resonant wavelengths in Fig.~\ref{fig:TPISpectra}~(c) as \mbox{$n_m = m\lambda_m/3L$}. We then performed a least-square fit to the $\beta$ vs. $\omega$ dispersion curve using a 4th-degree polynomial and computed the second-order derivative \mbox{$\beta_2 = d^2\beta/d\omega^2$} to obtain the GVD.\cite{Ferrera2009} Figure~\ref{fig:TPISpectra}~(c) plots the GVD curve in solid red, from which we obtain experimental GVD values of -495\textpm2, -535\textpm2, and -697\textpm2 ps$^2$/km at the signal, pump, and idler wavelengths,
respectively. For comparison, we also plotted in dashed red the simulated GVD of the FDMR loop obtained by averaging the GVDs of the SOI waveguide sections and the coupling regions comprising the loop. The simulated GVD of the FDMR is comparable to that of a straight SOI waveguide,\cite{Dulkeith2006} with GVD $\sim-4733$~ps$^2$/km around the pump wavelength of 1529 nm, which is larger than the experimental value. It has been demonstrated that GVD is strongly sensitive to cross-sectional area and aspect ratio of the SOI waveguide,\cite{Turner2006} and this is likely the source of discrepancy between the simulated and experimental values. However, the fact that the experimental GVD is less than or comparable to that of a straight waveguide over the 1510-1580 nm wavelength range demonstrates the low dispersion and broadband nature of the FDMR.
\par
\begin{figure}[tbp]
    \centering
    \includegraphics[width=0.48\textwidth]{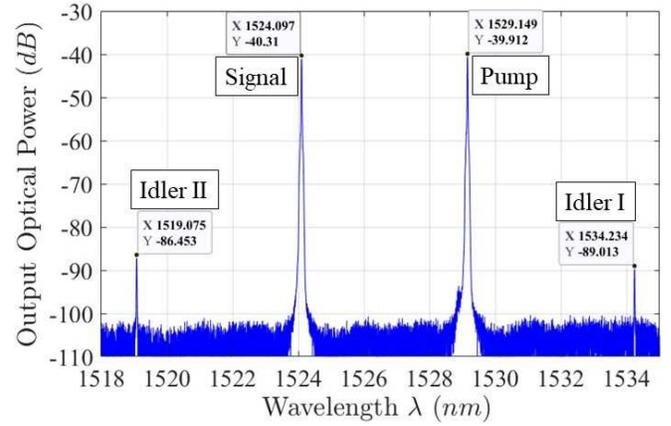}
    \caption{Transmission spectrum measured in the output waveguide of the Floquet microring lattice showing the generation of two idler waves by FWM.  The pump and signal were tuned to the linear FDMR resonance wavelengths.}
    \label{fig:FullFWMSpectra}
\end{figure}
\begin{figure*}[tbp]
    \centering
    \includegraphics[width=0.95\textwidth]{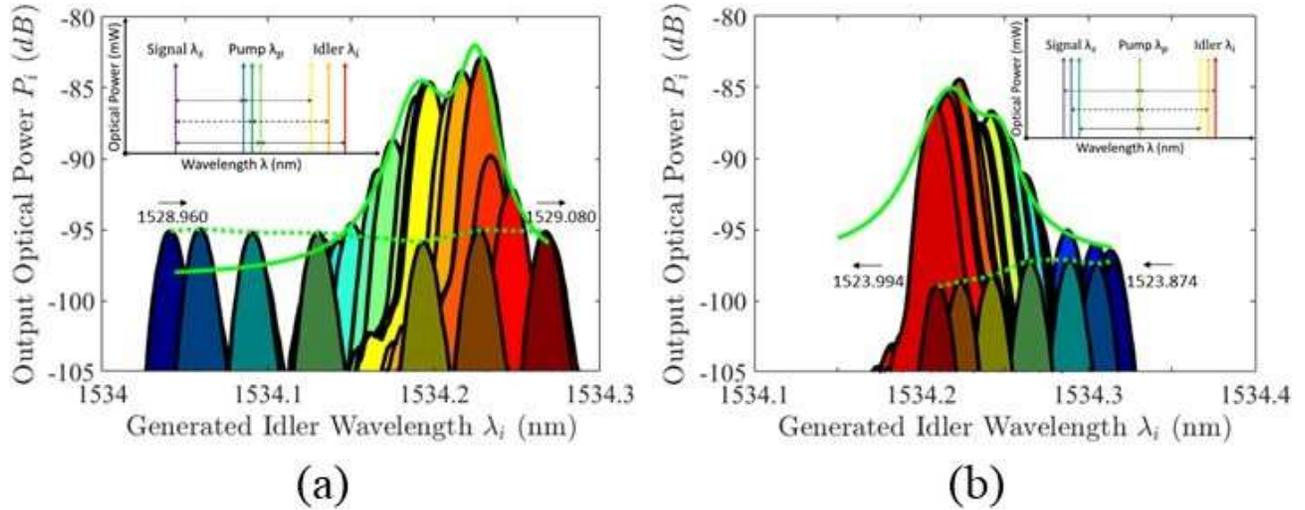}
        \caption{Spectra of output optical power at the idler wavelength. The darker foreground series are recorded without heating power applied showing FWM purely by edge modes; the lighter background series are recorded with a 17.8~mW heater power applied showing enhanced FWM by FDMR present. (a) $\lambda_p$ swept from blue (1528.960~nm) to red (1529.080~nm) wavelengths, $\lambda_i$ shifts from blue to red twice as quickly. (b) $\lambda_s$ swept from blue (1523.874~nm) to red (1523.994~nm) wavelengths, $\lambda_i$ shifts proportionally in reverse.}
        \label{fig:WavelengthRelationshipsAndResonance}
\end{figure*}

In the FWM experiment, with the heater power set at 17.8 mW, we tuned the signal and pump wavelengths to coincide with the FDMR at 1524.1~nm and 1529.1~nm, respectively. The spectral shapes of these resonances were similar, with Q-factor of 73,000 at the signal wavelength and 95,000 at the pump wavelength. However, at the 1534.2~nm resonance where the idler wave was generated, a split resonance spectrum was prominent, which was caused by coherent back-scattering into the counter-propagating mode in the FDMR loop and was not observed in neighbor resonances. As a result, we obtained a lower effective Q-factor of 36,000 for this resonance mode. By applying 2.55 mW of optical power for both the pump and signal into the input waveguide, we obtained the wavelength spectrum at the output waveguide in Fig.~\ref{fig:FullFWMSpectra}, which shows an idler generated at both 1519.1~nm and 1534.2~nm. However, due to a slight red shift in the FDMR resonances caused by the nonlinear thermo-optic effect at high optical powers (see supplementary material), the CE, defined as the ratio of the idler power to the signal power, was relatively low, around \mbox{-46~dB} and \mbox{-49~dB} for the up-converted and down-converted idler wavelengths, respectively. To optimize the CE, we swept the pump and signal wavelengths over the corresponding FDMR resonances to obtain the maximum generated idler power. Figure ~\ref{fig:WavelengthRelationshipsAndResonance} shows the variations in the generated idler power and wavelength as the pump wavelength is swept (Fig.~\ref{fig:WavelengthRelationshipsAndResonance}~(a)) and as the signal wavelength is swept (Fig.~\ref{fig:WavelengthRelationshipsAndResonance}~(b)). In each plot, the darker spectra in the foreground were recorded without the heating power applied, so that no FDMR was induced and the idler wave was generated purely from FWM of the edge modes propagating along the bottom lattice boundary. The lighter spectra in the background show the idler power enhancement resulting from the FDMR when the heating power was applied. The pump and signal wavelengths were sampled in 5~pm steps and the spectral shape of each idler spectrum corresponded to the line shapes of the input lasers.
\par
We observe that the idler power generated by edge mode FWM with the heating power turned off exhibits minimal wavelength dependence, which reflects the flat, wide-band transmission characteristics of the Floquet edge modes across the bandgaps. However, when the heating power was turned on, the peak optical power of the resonance-enhanced idler wave follows the shape of the resonance spectrum of the FDMR, approaching the edge mode results when either the pump or signal was tuned out of the resonance. The peak idler power generated due to FDMR is about 12 dB higher than due to the edge modes alone, and the 3dB bandwidth at which the idler power drops by a half is measured to be 51~pm and 45~pm for the pump- and signal-wavelength sweeps, respectively, which is in agreement with the linewidth of the linear idler FDMR spectrum.
\par
Since the generated idler wavelength is given by \mbox{$1/\lambda_i = 2/\lambda_p-1/\lambda_s$}, we obtain the rates of change of the idler wavelength with respect to the pump and signal wavelengths to be \mbox{$\delta\lambda_i/\delta\lambda_p = 2(\lambda_i/\lambda_p)^2$} and \mbox{$\delta\lambda_i/\delta\lambda_s = -(\lambda_i/\lambda_s)^2$}, indicating that the idler wavelength changes twice as fast as the pump wavelength changes and in the opposite direction as the signal wavelength changes. These trends are also confirmed by the wavelength sweeps in Figs.~\ref{fig:WavelengthRelationshipsAndResonance}~(a) and (b).  The wavelength conversion bandwidth, defined as the wavelength separation from the signal to the idler waves, is 10.1 nm and spans over 3 FB zones of the Floquet microring lattice. Such large conversion bandwidth is enabled by the low dispersion of the Floquet modes in the FDMR.
\begin{figure}[tbp]
         \centering
         \includegraphics[width=0.48\textwidth]{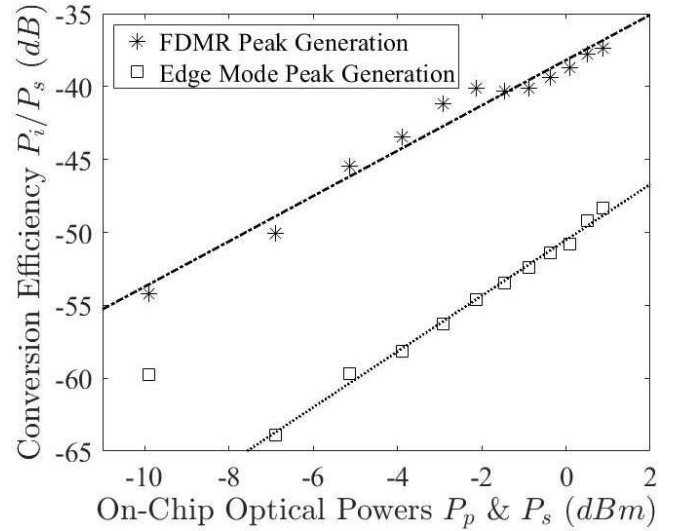}
     \caption{FWM CE versus on-chip optical power of pump and signal waves. The linear fits have slopes of 1.55 and 1.91 for the FDMR and edge mode, respectively. On average, the FDMR enhances the CE by 12.5 dB above that of the edge mode.}
     \label{fig:CE}
\end{figure}
\par
With the pump and signal wavelengths set at the 1529.1~nm and 1524.1~nm resonances, respectively, we swept the input optical powers of both the pump and signal from 102~$\mu$W to 1.23~mW in the input waveguide. At each input power setting, we optimized the spectral locations of the pump (1529.039~nm to 1529.121~nm) and signal (1523.929~nm to 1523.974~nm) wavelengths to maximize the idler power (1534.169~nm to 1534.227~nm). Figure~\ref{fig:CE} compares the CE of FWM enhanced by the FDMR and that due only to the edge modes, as a function of the pump (and signal) power. For comparison, the effective propagation length of the edge modes from input waveguide to output waveguide was 1119~$\mu$m, which is more than 3 times the effective circumference of the FDMR loop. For both cases, we observe that the CE increases with input powers,  with the FDMR providing a relatively constant average enhancement of 12.5~dB over the edge mode. At the highest available input pump power of 1.23~mW in the input waveguide, we obtained an internal CE of -37.4~dB with the FDMR present and -48.3~dB without. We also note that both CE trends do not show sign of levelling off at high powers, indicating that the detrimental effect of free carrier absorption (FCA) in the silicon microrings is negligible at these input power levels.
\section{Discussion and Conclusion}
The 330~$\mu$m roundtrip length of the FDMR gives an effective loop radius of 52.5$\mu$m, allowing our device to be directly compared to that of a 50~$\mu$m-radius SOI microring resonator.\cite{Turner2008} In their single-ring device, 2~mW of input pump power was required to reach a CE of -35~dB, with saturation effect due to FCA observed at 5~mW of input pump power. Projecting along the fitted curve in Fig.~\ref{fig:CE}, an input pump power of 5~mW would result in a resonance-enhanced CE of -27.3~dB in our FDMR. Our edge mode achieved -48.8~dB/mm at 1.23~mW pump power, which is comparable to the value of \mbox{-43~dB/mm} reported for a bare silicon waveguide\cite{Morichetti2011} but at a much higher pump power of 15.8~mW.
It has also been shown\cite{Morichetti2011} that by using the slow-light effect in a coupled-resonator optical waveguide (CROW) consisting of 8 microrings, a CE improvement of 28~dB over a bare waveguide could be achieved. The same approach can also be used to further increase the CE of our structure by coupling many FDMR loops together. It should be emphasized that while our FDMR provides comparable CE's to those reported for conventional silicon microrings, the main benefit of our device is that it enables light to be directly generated on a topologically protected platform.
\par
In summary, we report the first demonstration of resonance-enhanced FWM by localized bulk modes in a Floquet TPI lattice using a compact, cavity-less resonance effect induced through a periodic perturbation of the drive sequence. The strongly-localized FDMR, with tunable quasienergy across the TPI bandgap, provides high Q-factors and low dispersion, which enable efficient wavelength conversion over a broad bandwidth.  Combined with the topological protection of the Floquet edge modes, the system could provide a robust topological photonic platform for nonlinear optics applications such as wavelength conversion, parametric amplification, frequency comb and correlated photon pair generation.
\section*{Supplementary Material}
See supplementary material for an in-depth description of the Floquet-Bloch Hamiltonian of a 2D microring lattice with periodic perturbations, the experimental setup details, and a description of the nonlinear thermo-optic effect in FDMR at high optical powers.
\begin{acknowledgments}
This work is supported in part by financial contributions from the Natural Sciences and Engineering Research Council of Canada (CGSD3 - 558491 - 2021).
\end{acknowledgments}
\section*{Author Declarations}
\subsection*{Conflict of Interest}
The authors have no conflicts to disclose.
\section*{Data Availability Statement}
The data that supports the findings of this study are available within the article and its supplementary material.
\section*{References}
\bibliographystyle{apsrev4-1}
\bibliography{Bibliography}
\end{document}


\maketitle
\section{Floquet-Bloch Hamiltonian of a 2D Microring Lattice and Periodic Perturbations}
\begin{figure}[htbp]
         \centering
         \includegraphics[width=0.9\textwidth]{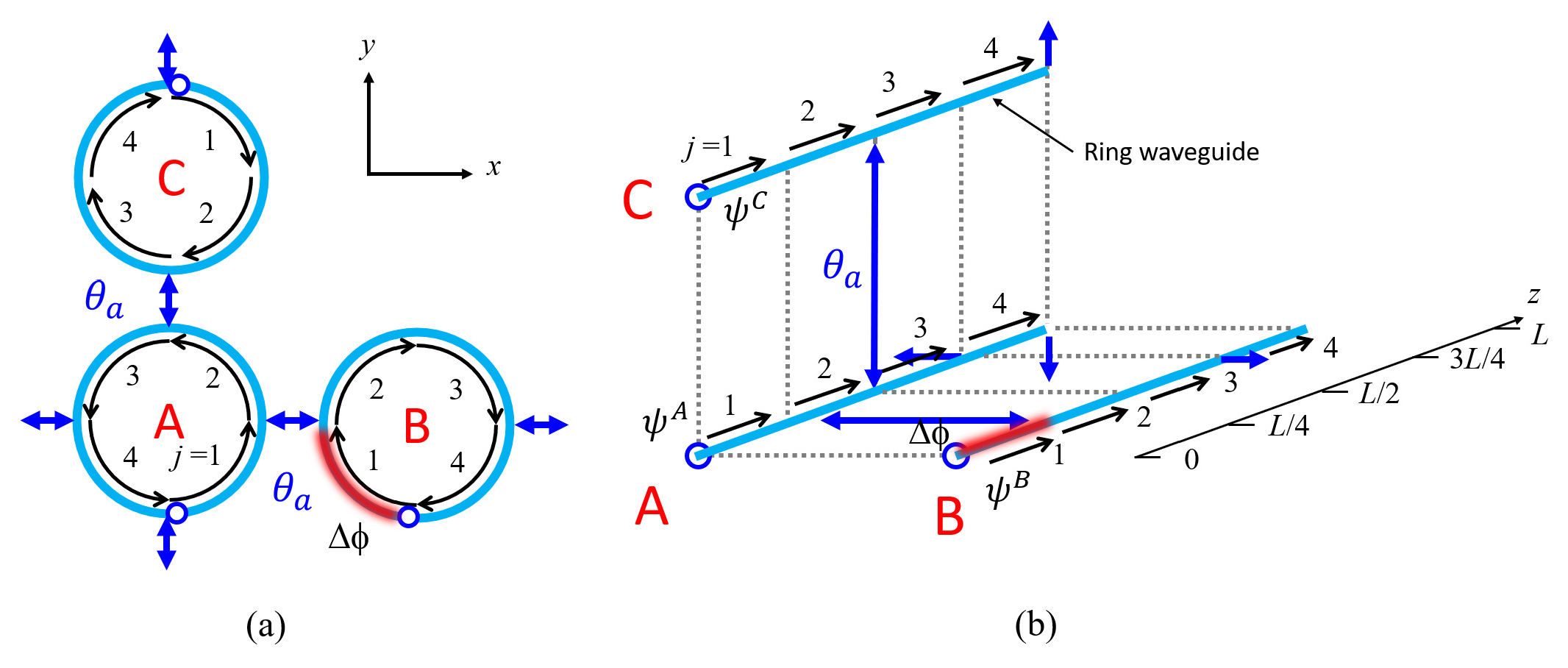}
     \caption{(a) Schematic of a unit cell of the 2D Floquet microring lattice with three microrings and neighbor coupling angle $\theta_a$. Each roundtrip of light propagation in a microring is broken into 4 steps, $j = 1$ to 4. Also shown is a periodic perturbation in the form of a phase detune $\Delta\phi$ (red) applied to step "$j=1$" of microring B. (b) Equivalent coupled waveguide array of each unit cell in the microring lattice. The field propagation is along the microring waveguides ("$z$"-axis) with each period equal to the microring circumference $L$. Waveguide B experiences a periodic phase detune of $\Delta\phi$ (red) in step "$j=1$".}
     \label{fig:Hamiltonian}
\end{figure}

The Floquet-Bloch Hamiltonian of our 2D microring lattice can be derived using a similar approach as for the general 2D microring lattice in \cite{Afzal2018}, except here each unit cell contains three microrings instead of four, as shown in Fig.~\ref{fig:Hamiltonian}~(a). We first transform the microring lattice into an equivalent 2D coupled waveguide array by "cutting" each microring in a unit cell at the points indicated by the small open circles in Fig.~\ref{fig:Hamiltonian}~(a) and unrolling it into a straight waveguide with length $L$ equal to the circumference of the microring, as shown in Fig.~\ref{fig:Hamiltonian}~(b). Thus, each roundtrip of light circulation in a microring is equivalent to light propagation along the coupled waveguide array of length $L$, which can be divided into a sequence of 4 coupling steps of equal length $L/4$. Denoting the fields in microrings A, B, and C in unit cell $(m,n)$ in the lattice as $\psi_{m,n}^A, \psi_{m,n}^B$ and $\psi_{m,n}^C$, respectively, we can write the equations for the evolution of light in the waveguide array along the direction of propagation ($z$-axis) over each period in terms of the Coupled Mode Equations
\begin{equation}
    -i\frac{\partial{\psi_{m,n}^A}}{\partial{z}} = \beta\psi_{m,n}^A + k_a(1)\psi_{m,n}^B + k_a(2)\psi_{m,n}^C + k_a(3)\psi_{m-1,n}^B + k_a(4)\psi_{m,n-1}^C
    \label{eq:S1}
\end{equation}
\begin{equation}
    -i\frac{\partial{\psi_{m,n}^B}}{\partial{z}} = \beta\psi_{m,n}^B + k_a(1)\psi_{m,n}^A + k_a(3)\psi_{m+1,n}^A 
    \label{eq:S2}
\end{equation}
\begin{equation}
    -i\frac{\partial{\psi_{m,n}^C}}{\partial{z}} = \beta\psi_{m,n}^C + k_a(2)\psi_{m,n}^A + k_a(4)\psi_{m,n+1}^A 
    \label{eq:S3}
\end{equation}
where $\beta$ is the propagation constant of the microring waveguide mode and $k_a(j)$ specifies the coupling between adjacent waveguides in step $j$ of each evolution period, with $k_a(j)=4\theta_a/L$ in step $j$ and $0$ otherwise.  By making use of the Bloch's boundary conditions $\psi_{m+1,n} = \psi_{m,n}e^{ik_x\Lambda}$, $\psi_{m,n+1} = \psi_{m,n}e^{ik_y\Lambda}$, where $\Lambda$ is the lattice constant, for each field in the microrings, we can express the Coupled Mode Equations in the form
\begin{equation}
-i\frac{\partial}{\partial z} |\psi(\textbf{k}, z)\rangle = [\beta I + H_{FB}(\textbf{k},z)]|\psi(\textbf{k}, z)\rangle 
\label{eq:S4}
\end{equation}
where $|\psi\rangle = [\psi_{m,n}^A,\psi_{m,n}^B,\psi_{m,n}^C]^T$, $\textbf{k}=(k_x,k_y)$ is the crystal momentum vector, and $I$ is the identity matrix.  The Floquet-Bloch Hamiltonian $H_{FB}$ consists of a sequence of 4 Hamiltonians $H(j)$ corresponding to coupling steps $j = \{1,2,3,4\}$, which have the explicit forms for $j = {1, 3}$ 
\begin{eqnarray*}
H(j) =& 
 \Bigg(
    \begin{matrix*}[c]
    0 & k_a e^{-i \delta_{3j} k_x \Lambda} & 0\\
    k_a e^{i \delta_{3j} k_x \Lambda} & 0 & 0\\
    0 & 0 & 0
    \end{matrix*}
    \ \Bigg)
\end{eqnarray*}
and for $j = {2, 4}$
\begin{eqnarray*}
H(j) =&
 \Bigg(\ 
    \begin{matrix*}[c]
    0 & 0 & k_a e^{-i \delta_{4j} k_y \Lambda}\\
    0 & 0 & 0\\
    k_a e^{i \delta_{4j} k_y \Lambda} & 0 & 0\\
    \end{matrix*}
    \Bigg) \\
\end{eqnarray*}
In the above expressions, $\delta_{ij} = 1$ if $i = j$ and 0 otherwise.  The Floquet-Bloch Hamiltonian is periodic in $z$ with a periodicity equal to the microring circumference $L$, $H_{FB}(\textbf{k},z)=H_{FB}(\textbf{k},z+L)$. The quasienergy spectrum of the lattice consists of three bands in each Floquet-Brillouin zone, with a flat band state at quasienergies $2m\pi,\  (m\in \mathbb{Z})$.  For $\theta_a \gtrsim \pi/\sqrt{8}$, all three bandgaps exhibit Anomalous Floquet Insulator behavior with nontrivial winding numbers \cite{Afzal2018}. 
\par
The FDMR is created by introducing a periodic perturbation to the Hamiltonian in the form of a phase detune of a microring during a step $j$ in each evolution period.  Thus, if a phase detune of $\Delta \phi$ is applied to microring B in a unit cell during step $j$, its equation of motion is modified as
\begin{equation}
    -i\frac{\partial{\psi_{m,n}^B}}{\partial{z}} = (\beta+ \Delta \beta(j))\psi_{m,n}^B + k_a(1)\psi_{m,n}^A + k_a(3)\psi_{m+1,n}^A 
    \label{eq:S5}
\end{equation}
where $\Delta \beta(j) = 4\Delta \phi /L$ in step $j$ and zero otherwise. We previously showed in \cite{AfzalAPL} that the spatial field localization pattern of the defect mode depends on which segments of the microring are detuned. To create an FDMR loop as shown in Fig.~1~(d) in the main text, one needs to apply the phase detune to segments $j = 1$ and $4$ of microring B (corresponding to the bottom half of the resonator).
\section{Experimental Setup}
\begin{figure}[htbp]
         \centering
         \includegraphics[width=0.75\textwidth]{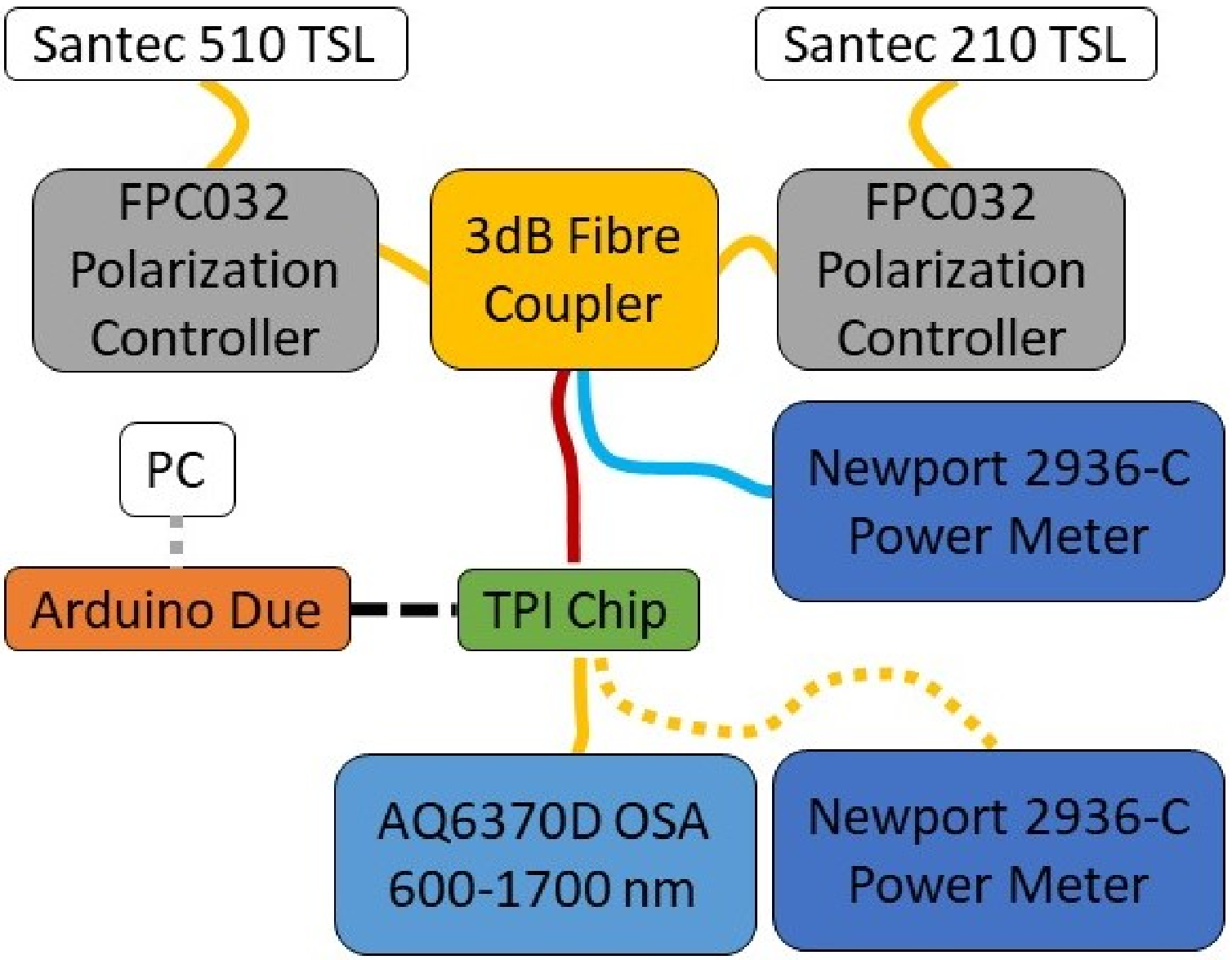}
     \caption{Experimental setup.  Two tunable semiconductor lasers passed through individual polarization controllers set to TE polarization. These signals were combined and measured by a power meter and sent to the chip (TPI Chip). The signal was butt-coupled into and out of the chip using lensed fibres. The output signal was measured by a power meter and the optical spectrum analyzer. The heating power applied to the defect ring was controlled by an Arduino Due board interfaced to a PC.}
     \label{fig:ExperimentalSetup}
\end{figure}

A schematic of the experimental setup is provided in Fig.~\ref{fig:ExperimentalSetup}. Two tunable laser sources (Santec 510 TSL and Santec 210 TSL) were employed, each coupled into a polarization controller (Thorlabs FPC032) set to TE polarization. The spectra of the lasers were nearly identical and maintained a constant spectral shape through various optical powers, with typical linewidth of 18~pm. When measuring the transmission spectrum of the TPI lattice, only the 510 TSL was used to probe the spectral response. the TE-polarized light was butt-coupled, using a lensed fibre, to the tapered edge of the SOI waveguide which served as the input port to the TPI. The transmitted light from the output waveguide was butt-coupled using another lensed-fiber and measured using a power meter (Newport 2936-C).
\par
For the FWM experiment, both TSL's were used to provide the pump and signal wavelengths and then combined using a 50/50 fiber coupler. The total loss from the laser output to the tip of the input lensed fibre was measured to be 5.9~dB. The combined pump and signal were then butt-coupled into the chip using a lensed fibre.  The output light was coupled to another lensed fibre and  measured using an optical spectrum analyzer (Yokogawa AQ6370D OSA 600 - 1700 nm). The fibre-to-waveguide coupling loss was calculated to be $\sim$4~dB per coupling.
\par
In the FWM experiment, the laser power was swept from 1 mW to 12 mW at the laser output, with both pump and signal wavelengths fine-tuned at each power level to find the FDMR resonant wavelengths. For each laser power, a heating current was applied to a circuit with a resistance R$\sim$300~$\Omega$ which was fabricated on top of the defect microring along the bottom edge of the lattice and accessed using metal pads fabricated away from the TPI lattice. A current of 7.7 mA was applied, resulting in 17.8 mW of heating power over $3/4$ of the defect ring circumference. This chosen current moved the FDMR resonant wavelengths into the center of stop band III, provided high Q-factor resonances, and resulted in fairly symmetric resonances across many FSRs that were used to demonstrate FWM.
\section{Nonlinear thermo-optic effect in the FDMR at high optical powers}
\begin{figure}[htbp]
         \centering
         \includegraphics[width=0.99\textwidth]{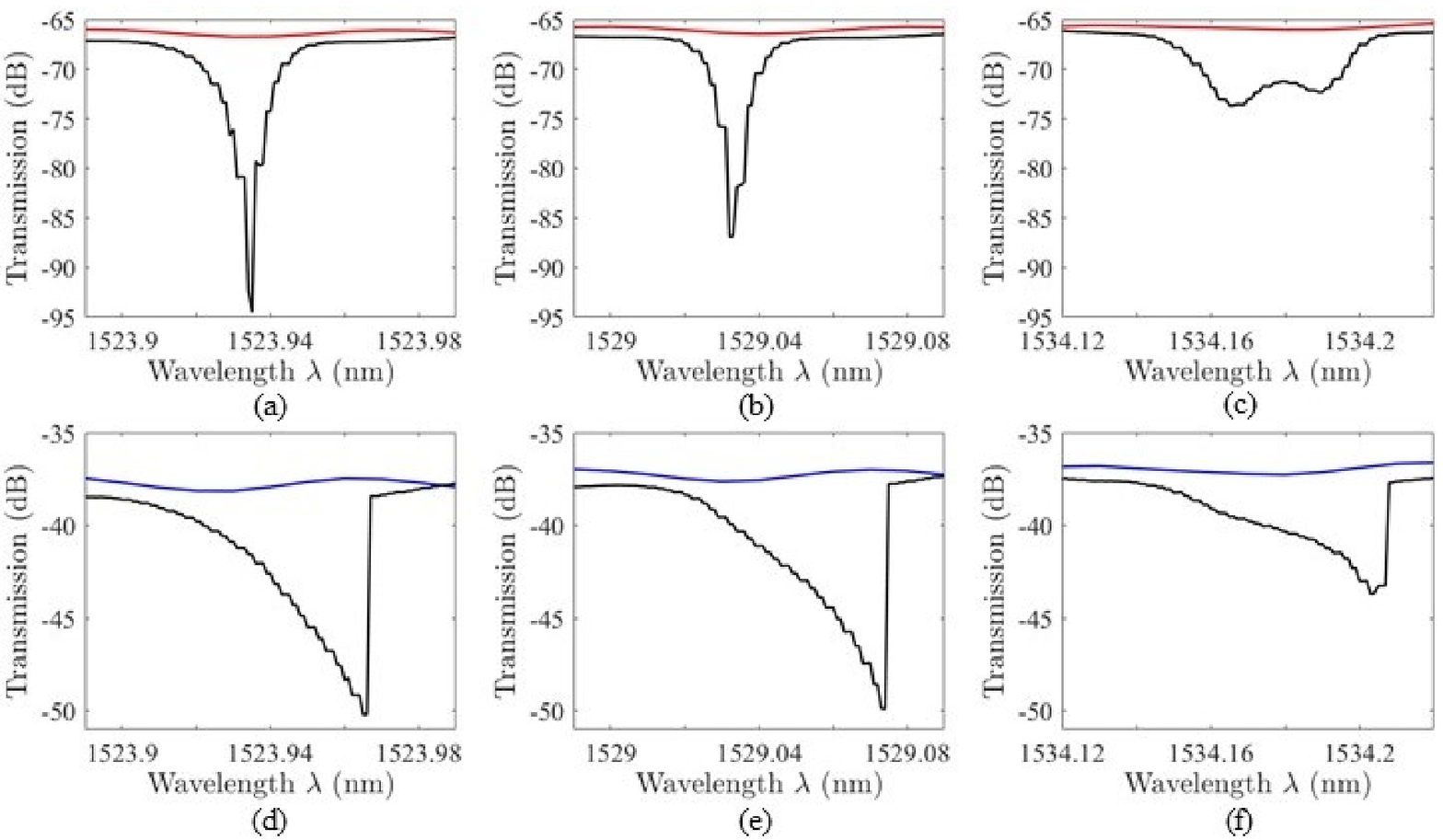}
     \caption{Spectral scans of the background TPI lattice without heating power applied are given in blue and red. The FDMR spectra with 3~$\mu$W input optical power are given in (a-c) and with 2.55~mW input optical power in (d-f).}
     \label{fig:NonlinearShift}
\end{figure}

At on-chip input laser powers $\gtrsim$ 0.5~mW, a red shift and skewing of the linear FDMR spectrum could be observed. This is due to the nonlinear thermo-optic effect in the silicon waveguide, which is caused by an intensity-dependent change in the refractive index of silicon. Figure~\ref{fig:NonlinearShift} demonstrates the difference in the measured spectra for input laser powers of 0.03~mW and 25~mW, which corresponded to 3~$\mu$W and 2.55~mW in the waveguide, respectively. Figures~\ref{fig:NonlinearShift}~(a), (b) and (c) show the resonance spectra at the signal, pump and idler wavelengths, respectively for 3~$\mu$W input power, while Figures~\ref{fig:NonlinearShift}~(d), (e) and (f) show the corresponding spectra at 2.55~mW input power. Due to the red shift and skewing of the spectra at high optical powers, it was necessary to fine tune the signal, pump and idler wavelengths in order to achieve maximum resonance enhancement for maximum wavelength conversion efficiency.  We also note that the double dip in the resonance spectrum at the idler wavelength was due to coherent back scattering.
\bibliography{Supplemental_Bibliography}